\documentstyle[12pt]{article}
\input amssym.def       
\input amssym.tex       

\def\goth{\frak}          
\def\double{\Bbb}

\def\cc{{\double C}}     
       
\def\rr{{\double R}}     
\def\zz{{\double Z}}

\def\mm{{{\cal M}}}

\def\ep{{\cal E}}

\def\ddd{{\,\hbox{$\partial\!\!\!/$}}}

\def\ot{\otimes}
\def\op{\oplus}

\def\mapright#1{\smash{\mathop{\longrightarrow}\limits^{#1}}}

\def\bb{\begin{eqnarray}}
\def\ee{\end{eqnarray}}
\def\eee{\nonumber\end{eqnarray}}

\newtheorem{definition}{Definition}[section]
\newtheorem{lemma}{Lemma}[section]

\begin{document}

\font\twelve=cmbx10 at 13pt
\font\eightrm=cmr8
\def\petit{\def\rm{\fam0\eightrm}}
\baselineskip 18pt

\begin{titlepage}
\title{(Fermionic)Mass Meets (Intrinsic)Curvature}
\author{J\"urgen Tolksdorf\thanks{email: tolkdorf@euler.math.uni-mannheim.de}\\
Inst. of Mathematics\\ University of Mannheim, Germany}
\date{Dec, 04 2002}
\maketitle

\begin{abstract}
Using the notion of vacuum pairs we show how the (square of the) 
mass matrix of the fermions can be considered geometrically as curvature. This 
curvature together with the curvature of space-time, defines the total curvature 
of the Clifford module bundle representing a ``free'' fermion within the 
geometrical setup of spontaneously broken Yang-Mills-Higgs gauge theories. 
The geometrical frame discussed here gives rise to a natural class of Lagrangian 
densities. It is shown that the geometry of the Clifford module bundle 
representing a free fermion is described by a canonical spectral invariant 
Lagrangian density. 
\end{abstract}
\end{titlepage}  

\section{Introduction}
In a recent paper we showed how bosonic mass is related to the extrinsic
geometry of a chosen vacuum (c.f. \cite{tolksdorf'02}). In the present paper we 
will show how the mass of a fermion is related to the curvature of the Hermitian
vector bundle that represents the (free) fermion in question. The geometrical
context we work with is that of Clifford module bundles and operators of Dirac
type. Using the notion of vacuum pairs we will show how the fermionic mass
matrix permits decomposition of the fermion bundle into the Whitney sum of 
certain Hermitian (line) bundles representing (almost) free fermions of specific
mass. A natural class of non-flat connection exists on this type of bundle which
is defined by the  space-time metric together with the mass of the fermion. The 
corresponding Dirac operator $\ddd_{\!\mbox{\tiny${\cal D}$}}$ is the geometrical 
analogue of Dirac's first order operator $i\ddd - m$ that has been introduced
to relativistically describe the dynamics of a free fermion of mass m. We show
how a (linear) fluctuation of the vacuum yields the Dirac-type operator 
$D_{\mbox{\tiny Y}}$ usually referred to as Dirac-Yukawa operator.\\
 
The basic question addressed in this paper is how to understand the notion
of fermionic mass from a geometrical perspective. Interestingly, this question
is tied to several other basic questions like, for instance, how to understand 
Dirac's famous first order operator as a ``true'' Dirac operator. Or, since the 
notion of mass is related to the notion of a ``free'' particle (i.e. to a dynamically 
closed system\footnote{It is well-known that, e.g., in the context of the 
strong interaction there is no unique definition of mass of the quarks.}), 
how one can 
understand the notion of ``freeness'' within the geometrical setup of Yang-Mills
gauge theories. Another question along this line of thought is how one can 
geometrically understand what is usually referred to as ``particle multiplet''.
Usually, elementary particles are described by (``quantized'') fields $\Psi$ that 
are defined on space-time $\mm$ without referring to a geometrical description 
of the particles themselves. Moreover, it is assumed that some of these fields 
actually constitute a ``fermion multiplet'' with respect to some ``internal 
symmetry group'' G. As is well-known, for instance, in the case of the Standard 
Model of particle physics the fields representing a (left handed) electron and a 
(left handed) neutrino together build a (left handed) fermion doublet with respect 
to the symmetry group ${\rm SU(2)}\times{\rm U(1)}$ of the electroweak 
interaction. Thus, if we believe in the Standard Model, neither an electron nor a 
neutrino itself can actually be regarded as a fundamental particle. Moreover, 
electromagnetism itself turns out to be an effective interaction only. 
Consequently, the field $\Psi$ describing the fermion doublet, e.g, of an electron 
and a neutrino is considered to decompose into $\Psi=(\Psi_1,\Psi_2)$. The gauge
symmetry of the electroweak interaction then manifests itselves in the 
arbitrariness of which component of $\Psi$ is identified with the field 
describing, e.g., an electron. In other words, one usually has to choose a
gauge in order to identify, for instance, $\Psi_1$ with the electron field.
However, such a description seems to be unsatisfying since on the one hand
the choice of a gauge is a purely mathematical operation (i.e. it cannot be 
achieved experimentally). On the other hand, there is no doubt that an 
electron exists in nature as an object of its own. It thus cannot depend on some 
choice of gauge.\\

From a purely mathematical point of view the space where $\Psi$ takes its
values forms a specific representation of the symmetry group G in question.
For instance, in the case of the electroweak group this space\footnote{Here, 
only one generation of left handed fermions is taken into account in order to 
simplify the discussion. The general case is discussed below.} is identified with
$\cc^{\mbox{\tiny 2}}$. Then, the fermionic mass matrix ${\rm M}_{\mbox{\tiny F}}$ 
provides a natural decomposition of this space into the eigenspaces of the mass 
matrix. In the case of ${\rm G}={\rm SU(2)}\times{\rm U(1)}$ one then obtains
\bb
\label{fermiondecomp}
\cc^{\mbox{\tiny 2}}\simeq 
W_{\mbox{\tiny electron}}\op W_{\mbox{\tiny neutrino}}.
\ee

This decomposition breaks the original gauge symmetry since the fermionic 
mass matrix ${\rm M}_{\mbox{\tiny F}}$ does not, in general, lie in 
the commutant of the symmetry group G. However, the point is that the
decomposition (\ref{fermiondecomp}) does not refer to any specific gauge.
The decomposition is ``natural'' with respect to the additional piece of input 
that comes from the fermionic mass matrix. However, in order to put the 
decomposition (\ref{fermiondecomp}) in an appropriate geometrical context 
without assuming the triviality of the underlying gauge bundle we first have 
to globalize the fermionic mass matrix. This will be done by using vacuum pairs
similar to the case of the bosonic mass matrices (see loc sit). 
As a consequence 
we will see that the mass matrix has a simple geometrical interpretation in 
terms of curvature and that Dirac's operator can in fact be considered as a 
Dirac-type operator.

\section{Orbit Bundles and Vacuum Pairs}
To get started we first summarize the notion of vacuum pairs that has been
introduced in \cite{tolksdorf'02}. For this let ${\cal P}(\mm,{\rm G})$ be a
smooth principal G-bundle ${\rm P}\mapright{\pi_{\rm P}}\mm$ over a smooth
orientable (pseudo)Riemannian (spin-) manifold $(\mm,g_{\mbox{\tiny M}})$ of 
dimension ${\rm dim}(\mm)=2n$. Here, G is a semi-simple compact real Lie group
with Lie algebra Lie(G). The corresponding gauge group is denoted by ${\cal G}$. Let 
${\rm G}\mapright{\rho_{\rm H}}{\rm Aut}(\cc^{\mbox{\tiny${\rm N}_{\rm H}$}})$
be a unitary representation of G. Also, let 
$\cc^{\mbox{\tiny${\rm N}_{\rm H}$}}\mapright{V_{\rm H}}\rr$ be a smooth
G-invariant function that is bounded from below. Moreover, it is assumed that 
its Hessian is positive definite transversally to the orbits of minima. We call 
$V_{\mbox{\tiny H}}$ a {\it general Higgs potential}. The triple
$({\cal P}(\mm,{\rm G}),\rho_{\mbox{\tiny H}},V_{\mbox{\tiny H}})$ defines the
geometrical data of a {\it Yang-Mills-Higgs gauge theory}. We call the Hermitian 
vector bundle $\xi_{\mbox{\tiny H}}$:
\bb
\pi_{\mbox{\tiny H}}:\,
{\rm E}_{\mbox{\tiny H}}:={\rm P}\times_{\rho_{\mbox{\tiny H}}}
\cc^{\mbox{\tiny${\rm N}_{\rm H}$}}&\longrightarrow&\mm,\cr
{\goth Z}=[(p,{\bf z})]&\mapsto&\pi(p),
\ee
the {\it Higgs-bundle} with respect to the above given data. It is assumed to
geometrically represent the Higgs boson. Correspondingly, a state of the
Higgs boson is geometrically represented by a section of the Higgs bundle.\\

Each minimum ${\bf z}_0\in\cc^{\mbox{\tiny${\rm N}_{\rm H}$}}$ defines a
sub-bundle of the Higgs bundle. For this, let  
${\rm orbit}({\bf z}_0)\subset\cc^{\mbox{\tiny${\rm N}_{\rm H}$}}$ and 
${\rm I}({\bf z}_0)\subset{\rm G}$ be the orbit and the isotropie group of the 
minimum. We call the fiber bundle $\xi_{\mbox{\tiny${\rm orbit}({\bf z}_0)$}}$:
\bb
\pi_{\mbox{\tiny orb}}:\,
{\cal O}rbit({\bf z}_0):={\rm P}\times_{\rho_{\rm orb}}{\rm orbit}({\bf z}_0)
\longrightarrow\mm
\ee
the {\it orbit bundle} with respect to the data defining a Yang-Mills-Higgs
gauge theory (see above). Here, $\rho_{\mbox{\tiny orb}}:=
\rho_{\mbox{\tiny H}}|_{\mbox{\tiny${\rm orbit}({\bf z}_0)$}}$.\\

Notice that, since 
$\xi_{\mbox{\tiny${\rm orbit}({\bf z}_0)$}}\subset\xi_{\mbox{\tiny H}}$,
every section ${\cal V}\in\Gamma(\xi_{\mbox{\tiny${\rm orbit}({\bf z}_0)$}})$
of the orbit bundle can be also considered as a section of the Higgs bundle.
There is a one-to-one correspondence between the sections ${\cal V}$ and
``H-reductions'' of ${\cal P}(\mm,{\rm G})$, where ${\rm H}\simeq{\rm I}({\bf z}_0)$.
More precisely, let H be the unique subgroup of G that is similar to the
isotropie group of the minimum ${\bf z}_0$. Then, every section
${\cal V}\in\xi_{\mbox{\tiny${\rm orbit}({\bf z}_0)$}}$ uniquely corresponds to
a principal H-bundle ${\cal Q}(\mm,{\rm H})$ together with an embedding 
${\rm Q}\mapright{\iota}{\rm P}$, such that the following 
diagram commutes (see, for instance, Ch. 1, Prop. 5.6 in \cite{koba/nomi})
\begin{center}
\setlength{\unitlength}{1.0cm}
\begin{picture}(3,3)
\put(-0.18,1.3){$\pi_{\mbox{\tiny Q}}$}
\put(1.0,1.3){$\pi_{\mbox{\tiny P}}$}
\put(1.85,1.3){$\pi_{\mbox{\tiny P}}$}
\put(1.25,0.25){$\pi_{\mbox{\tiny${\rm orb}$}}$}
\put(2.8,1.3){$\kappa$}
\put(1.4,2.6){$\iota$}
\put(0,0){$\mm$}
\put(2.40,0.0){${\cal O}rbit({\bf z}_0)$}
\put(0.15,2.4){${\rm Q}$}
\put(2.6,2.4){${\rm P}$}
\thicklines\put(0.32,2.3){\vector(0,-1){2}}
\thicklines\put(2.32,0.1){\vector(-1,0){1.8}}
\thicklines\put(2.7,2.3){\vector(0,-1){2}}
\thicklines\put(0.6,2.5){\vector(1,0){1.8}}
\thicklines\put(2.6,2.3){\vector(-1,-1){2.1}}
\end{picture}
\end{center}

Note that ${\rm P}\mapright{\kappa}{\cal O}rbit({\bf z}_0)$ is a principal 
H-bundle, 
where $\kappa(pg):=[(p,\rho(g){\bf z}_0)]$ denotes the canonical projection.\\

We call a section ${\cal V}$ of the orbit bundle a {\it vacuum section} and 
$({\cal Q},\iota)$ the corresponding {\it vacuum} with respect to the minimum 
${\bf z}_0$. We denote by ${\cal H}$ the gauge group that is defined by the
vacuum $({\cal Q},\iota)$ and call it the {\it invariance group} of the vacuum. 
A Yang-Mills-Higgs gauge theory is called {\it spontaneously broken} by a
vacuum $({\cal Q},\iota)$ if the invariance group of the latter is a proper
subgroup of the original gauge group ${\cal G}$. The gauge theory is called
{\it completely broken} by the vacuum if the appropriate invariance group 
is trivial. We call a vacuum $({\cal Q},\iota)$ trivial if ${\cal Q}(\mm,{\rm H})$ 
is equivalent to the trivial principal H-bundle 
$\mm\times{\rm H}\mapright{{\rm pr}_1}\mm$. Notice that even a trivial 
gauge bundle ${\cal P}(\mm,{\rm G})$ may have nontrivial vacua.\\

A connection $A$ on ${\cal P}(\mm,{\rm G})$ is called {\it reducible} with
respect to a given vacuum $({\cal Q},\iota)$ (or {\it compatible} with respect to the
vacuum section ${\cal V}$) if $\iota^*A$ is a connection on ${\cal Q}(\mm,{\rm H})$.
Let, respectively, ${\cal A}(\xi_{\mbox{\tiny H}})$ and
$\Gamma(\xi_{\mbox{\tiny H}})$ be the affine set of all associated connections on
the Higgs bundle and the module of sections of the Higgs bundle. A 
Yang-Mills-Higgs pair $(\Theta,{\cal V})\in
{\cal A}(\xi_{\mbox{\tiny H}})\times\Gamma(\xi_{\mbox{\tiny H}})$ is called
a {\it vacuum pair} if ${\cal V}$ is a vacuum section and
$\Theta$ corresponds to a flat connection on ${\cal P}(\mm,{\rm G})$ that is
compatible with the vacuum section.
Clearly, a vacuum pair minimizes the energy functional that corresponds to the 
Yang-Mills-Higgs action with respect to the data 
$({\cal P}(\mm,{\rm G}),\rho_{\mbox{\tiny H}},V_{\mbox{\tiny H}})$. In particular,
a vacuum section ${\cal V}$ corresponds to a ground state of the Higgs boson.
A vacuum pair $(\Theta,{\cal V})$ geometrically generalizes the canonical 
vacuum pair $(d,{\bf z}_0)$ in the case of the trivial gauge bundle
$\mm\times{\rm G}\mapright{{\rm pr}_1}\mm$. In fact, it can be shown that
in the case of a simply connected space-time there is at most one vacuum pair 
to the orbit of a given minimum ${\bf z}_0$ apart from gauge equivalence. 
Moreover, this pair is gauge equivalent to the canonical vacuum 
pair\footnote{Here, a minimum ${\bf z}_0$ is considered as the vacuum section
$\mm\mapright{{\bf z}_0}\mm\times{\rm orbit}({\bf z}_0),\,x\mapsto(x,{\bf z}_0)$
and $d$ is the covariant derivative with respect to the trivial connection. 
Indeed, in physics the mechanism of spontaneous symmetry breaking refers to
the canonical vacuum pair $(d,{\bf z}_0)$. This is consistent for in particle physics
the common model of space-time is that of  
$(\mm,g_{\mbox{\tiny M}})\simeq\rr^{\mbox{\tiny 1,3}}$.}. In general, however, 
the data 
$({\cal P}(\mm,{\rm G}),\rho_{\mbox{\tiny H}},V_{\mbox{\tiny H}})$ may give
rise to gauge inequivalent vacua even in the case of only one nontrivial orbit 
of minima.\\

We have summarized the basic geometrical notion that we need to globalize
the fermionic mass matrix. This will be discussed in the next section.

\section{Clifford Module Bundles and the Fermionic Mass Matrix}
Having chosen a spin structure ${\cal S}$ we denote the appropriate spinor 
bundle by $\xi_{\mbox{\tiny S}}$. Let 
${\rm G}\mapright{\rho_{\rm F}}{\rm Aut}(\cc^{\mbox{\tiny${\rm N}_{\rm F}$}})$
denote a second unitary representation of G. The corresponding associated
Hermitian vector bundle $\zeta_{\mbox{\tiny F}}$ is defined by:
\bb
\pi_{\mbox{\tiny F}}:\,{\rm E}_{\mbox{\tiny F}}:=
{\rm P}\times_{\rho_{\rm F}}\cc^{\mbox{\tiny${\rm N}_{\rm F}$}}
\longrightarrow\mm.
\ee
We then call the twisted spinor bundle 
\bb
\xi_{\mbox{\tiny F}}:=\xi_{\mbox{\tiny S}}\ot\zeta_{\mbox{\tiny F}}
\ee
the {\it fermion bundle} with respect to the data
$({\cal P}(\mm,{\rm G}),\rho_{\mbox{\tiny F}},{\cal S})$. It geometrically represents 
a particle of spin one-half. In what follows we will assume that the fermion bundle 
is also $\zz_2-$graded with respect to the ``inner degrees of freedom'', i.e.
$\zeta_{\mbox{\tiny F}}=\zeta_{\mbox{\tiny F,L}}\op\zeta_{\mbox{\tiny F,R}}$.\\

On such a $\zz_2-$graded fermion bundle there exists a distinguished class of
first order differential operators called {\it Dirac operators of simple type}
(c.f. \cite{ackerm/tolksd'96} and \cite{tolksdorf'98}). More precisely, let
$\xi_{\mbox{\tiny Cl}}$ be the Clifford bundle with respect to 
$(\mm,g_{\mbox{\tiny M}})$. The fermion bundle forms a natural left module of
the Clifford bundle. The corresponding action is denoted by $\gamma$. By an
operator of Dirac type we mean any odd first order differential operator $D$
acting on the module of sections $\Gamma(\xi_{\mbox{\tiny F}})$ such that
$D^2$ is a generalized Laplacian, i.e. $[D,[D,f]]=\pm2 g_{\mbox{\tiny M}}(df,df)$
for all $f\in{\cal C}^\infty(\mm)$ (see, e.g., Ch. 3.3 in \cite{BGV}). Let 
${\cal D}(\xi_{\mbox{\tiny F}})$ be the affine set of all operators of Dirac type 
which are compatible with the Clifford action $\gamma$, i.e. $[D,f]=\gamma(df)$.
The appropriate vector space is given by $\Omega^0(\mm,{\rm End}^-(\ep))$,
where $\ep:={\rm S}\ot{\rm E}_{\mbox{\tiny F}}$ is the total space of the fermion
bundle. We also denote by ${\cal A}(\xi_{\mbox{\tiny F}})$ the affine set of all
(associated) connections on $\xi_{\mbox{\tiny F}}$. The corresponding vector
space is given by $\Omega^1(\mm,{\rm End}^+(\ep))$. In general, one has
${\cal D}(\xi_{\mbox{\tiny F}})\simeq
{\cal A}(\xi_{\mbox{\tiny F}})/{\rm Ker}(\gamma)$. Thus, there is a whole
class $[A]$ of connections on the fermion bundle corresponding to each Dirac
type operator $D$. However, there is a distinguished class of connections on 
the fermion bundle that is constructed as follows: An operator
$D\in{\cal D}(\xi_{\mbox{\tiny F}})$ is called of simple type if its Bochner-Laplacian
$\triangle_{\mbox{\tiny D}}$ is defined by a Clifford connection
$A\in{\cal A}_{\mbox{\tiny Cl}}(\xi_{\mbox{\tiny F}})\subset
{\cal A}(\xi_{\mbox{\tiny F}})$. Here,  
${\cal A}_{\mbox{\tiny Cl}}(\xi_{\mbox{\tiny F}})$ denotes
the affine subset of Clifford connections on the fermion bundle. They are
characterized by the covariant derivatives $\partial_{\!\mbox{\tiny A}}$ that 
fulfill $[\partial_{\!\mbox{\tiny A},X}\, ,\gamma(a)]=
\gamma(\partial^{\mbox{\tiny Cl}}_{\! X}a)$
for all sections $a\in\Gamma(\xi_{\mbox{\tiny Cl}})$ and tangent vector fields 
$X\in\Gamma(\tau_{\mbox{\tiny M}})$. Here, $\partial^{\mbox{\tiny Cl}}$
is the covariant derivative that is defined by the canonical connection on the 
Clifford bundle $\xi_{\mbox{\tiny Cl}}$. In the case of a twisted spinor bundle
Clifford connections are tensor product connections and thus are parameterized
by Yang-Mills connections on $\zeta_{\mbox{\tiny F}}$. It can be shown that 
$D$ is of simple type iff it reads (c.f. \cite{ackerm/tolksd'96} and
\cite{tolksdorf'98})
\bb
\label{dop of simple type}
D\equiv\ddd_{\!\mbox{\tiny${\rm A},\phi$}} = 
\ddd_{\!\mbox{\tiny A}} + \gamma_5\ot\phi,
\ee
where $\gamma_5$ is the grading operator on $\xi_{\mbox{\tiny S}}$ and
$\phi\in\Omega^0(\mm,{\rm End}^-({\rm E}_{\mbox{\tiny F}}))$.\\ 

Of course, any twisted Spin-Dirac operator 
$\ddd_{\!\mbox{\tiny A}}:=\gamma\circ\partial_{\!\mbox{\tiny A}}$
is of simple type. However, the most general Dirac operator of simple
type on the fermion bundle (more general: on any ``twisted'' Clifford module 
bundle) is given by (\ref{dop of simple type}). Notice that these more general
Dirac operators exist only if $\zeta_{\mbox{\tiny F}}$ is $\zz_2-$graded. The
connection class of a Dirac operator of simple type has a natural representative.
The corresponding covariant derivative reads
\bb
\partial_{\!\mbox{\tiny${\rm A},\phi$}} = 
\partial_{\!\mbox{\tiny A}} + \xi\wedge(\gamma_5\ot\phi).
\ee
Here, $\xi\in\Omega^1(\mm,{\rm End}^-(\ep))$ is the canonical one form
that fulfills the following criteria: a) it is covariantly constant with respect 
to every Clifford connection; b) it defines a right inverse of the Clifford action 
$\gamma$ (c.f \cite{tolksdorf'98}).\\

\begin{definition}
Let $({\cal P}(\mm,{\rm G}),\rho_{\mbox{\tiny H}},V_{\mbox{\tiny H}})$ be
the data of a Yang-Mills-Higgs gauge theory and let $\xi_{\mbox{\tiny F}}$ 
be the fermion bundle with respect to
$({\cal P}(\mm,{\rm G}),\rho_{\mbox{\tiny F}},{\cal S})$. A linear mapping
\bb
\label{yukawa mapping}
G_{\mbox{\tiny Y}}: \Gamma(\xi_{\mbox{\tiny H}})&\longrightarrow&
\Gamma(\xi_{\mbox{\tiny${\rm End}^-({\rm E}_{\rm F})$}})\cr
\varphi&\mapsto&\phi_{\mbox{\tiny Y}}:=G_{\mbox{\tiny Y}}(\varphi),
\ee 
such that $G_{\mbox{\tiny Y}}(\varphi)^\dagger=-G_{\mbox{\tiny Y}}(\varphi)$
is called a ``Yukawa mapping''. A Dirac operator of simple type
\bb
\label{dirac yukawa operator}
D_{\mbox{\tiny Y}}:=\ddd_{\!\mbox{\tiny A}}+\gamma_5\ot\phi_{\mbox{\tiny Y}}
\ee
is called a ``general Dirac-Yukawa operator''. Moreover, if $({\cal Q},\iota)$ is
a vacuum that spontaneously breaks the gauge symmetry, then the 
Hermitian section 
\bb 
-i{\cal D} &:=& -iG_{\mbox{\tiny Y}}({\cal V})\\[0.15cm]
&\equiv& 
{\mbox{\small$\left(\begin{array}{cc}
          0                                                &   {\rm M}_{\mbox{\tiny F}} \cr
{\rm M}_{\mbox{\tiny F}}^\dagger &         0       \end{array}\right)$}}
\ee 
is called the ``fermionic mass matrix''.
\end{definition} 

Clearly, a necessary condition for the existence of a Yukawa mapping is that
the representation $\rho_{\mbox{\tiny H}}$ and the fermionic representation 
$\rho_{\mbox{\tiny F}}$ are not independent of each other. For instance, in the 
case of the (minimal) Standard Model the existence of (\ref{yukawa mapping}) 
is equivalent to the validity of the well-known relations between the 
``hyper-charges'' of the leptons, the quarks and the Higgs boson
(see, e.g., \cite{tolksdorf'98}). The constants which parameterize the mapping
$G_{\mbox{\tiny Y}}$ are usually referred to as ``Yukawa coupling constants''.\\

In this section we have seen how the notion of vacuum (pairs) can be used
to consider the fermionic mass matrix as a globally defined (odd) operator 
acting on the states of a fermion that is geometrically defined by the data 
$({\cal P}(\mm,{\rm G}),\rho_{\mbox{\tiny F}},{\cal S})$. In the next section
we will show how the fermionic mass matrix ${\cal D}$ (together with the 
(pseudo) metric $g_{\mbox{\tiny M}}$) defines a canonical connection on the 
``reduced'' fermion bundle. A necessary condition for this connection to be 
flat 
is that the (almost) ``free fermions'' are massless. Moreover, the fermionic mass 
matrix will provide us with a geometrical interpretation of the ``minimal coupling'' 
in terms of the physically intuitive notion of ``fluctuating vacua''. The main feature 
of this geometrical interpretation is that, besides the Yang-Mills boson, the 
minimal coupling naturally includes the gravitational field and the Higgs boson.

\section{Dirac-Yukawa Operators as Fluctuating Vacua}
The Yukawa mapping (\ref{yukawa mapping}) permits us to consider a section of
the Higgs bundle (i.e. a state of the Higgs boson) as an (odd) endomorphism 
acting on the fermion bundle. In particular, a vacuum pair $(\Theta,{\cal V})$ 
defines a Dirac-Yukawa operator
\bb
\label{vacuum dop}
\ddd_{\!\mbox{\tiny${\cal D}$}} := \ddd + \gamma_5\ot{\cal D}
\ee
acting on sections of the {\it reduced fermion bundle} 
$\xi_{\mbox{\tiny F,red}}:=\xi_{\mbox{\tiny S}}\ot\zeta_{\mbox{\tiny F,red}}$,
with $\zeta_{\mbox{\tiny F,red}}$ defined by
\bb
\pi_{\mbox{\tiny F,red}}:\,{\rm E}_{\mbox{\tiny F,red}}:=
{\rm Q}\times_{\rho_{\rm F,red}}\cc^{\mbox{\tiny${\rm N}_{\rm F}$}}
&\longrightarrow&\mm\cr
{\goth Z}=[(q,{\bf z})]&\mapsto&\pi_{\mbox{\tiny Q}}(q).
\ee
Here, $\rho_{\mbox{\tiny F,red}}:=\rho_{\mbox{\tiny F}}|_{\mbox{\tiny H}}$.
Notice that $\xi_{\mbox{\tiny F,red}}\simeq\xi_{\mbox{\tiny F}}$. Thus, a section
of the reduced fermion bundle can be considered as a state of the fermion that
refers to a particular vacuum (pair)\footnote{This is analogous to the reduced
tangent bundle of an O(2n)-reduction of the frame bundle of $\mm$: A local
frame corresponds to 2n locally linear independent sections of the tangent
bundle $\tau_{\mbox{\tiny M}}$ that are orthonormal with respect to the
chosen reduction.}.\\

As a consequence, a vacuum pair defines a natural non-flat connection on the 
reduced fermion bundle. This connection is defined by the 
{\it Dirac-Yukawa operator in the vacuum state} (\ref{vacuum dop}). The 
appropriate covariant derivative reads
\bb
\partial_{\!\mbox{\tiny${\cal D}$}} := \partial + \xi\wedge(\gamma_5\ot{\cal D}).
\ee

The (total) curvature on $\xi_{\mbox{\tiny F,red}}$, which is defined by 
$(g_{\mbox{\tiny M}},\Theta,{\cal V})$, is given by\footnote{We would like to
point out that all of this can also be defined without assuming the existence of
a spin structure. Thus, it is the gravitational field together with the vacuum 
(pair) that counts and not so much the spin structure ${\cal S}$. At least, this
holds true as long as the notion of anti-particles is not taken into account.}
\bb
{\cal F}_{\!\!\mbox{\tiny${\cal D}$}} = 
{/\!\!\!\!{\cal R}} + {\rm m}_{\mbox{\tiny F}}^2\,\xi\wedge\xi,
\ee
where ${/\!\!\!\!{\cal R}}$ denotes the lifted (pseudo)Riemannian curvature tensor
with respect to $g_{\mbox{\tiny M}}$ and 
$i{\rm m}_{\mbox{\tiny F}}:=\gamma_5\ot{\cal D}$.\\

The relative curvature ${\cal F}_{\!\!\mbox{\tiny${\cal D}$}}^{\mbox{\tiny$\ep/S$}}
={\rm m}_{\mbox{\tiny F}}^2\,\xi\wedge\xi$ on the reduced fermion bundle is 
thus defined by the (square of the) mass matrix of the fermion with respect to 
the chosen vacuum (pair). Like in the case of the bosonic mass matrices we 
have the following

\begin{lemma}
The spectrum of the fermionic mass matrix is constant and only depends on 
the orbit of the minimum ${\bf z}_0$ of a general Higgs potential. Moreover, 
the mass matrix lies within the commutant of the invariance group of the 
vacuum chosen. Hence, the reduced fermion bundle splits into the 
Whitney sum of the eigenbundles of the fermionic mass matrix, i.e.
\bb
\xi_{\mbox{\tiny F,red}}\; =
\bigoplus_{\mbox{\tiny$m^2\in{\rm spec}({\rm m}_{\rm F}^2)$}}
\xi_{\mbox{\tiny F}}{\mbox{\tiny$({\rm m}^2)$}},
\ee 
where $\xi_{\mbox{\tiny F}}{\mbox{\tiny$({\rm m}^2)$}} :=
\xi_{\mbox{\tiny F,L}}{\mbox{\tiny$({\rm m}^2)$}}\op
\xi_{\mbox{\tiny F,R}}{\mbox{\tiny$({\rm m}^2)$}}$.
\end{lemma}

\noindent
{\bf Proof:} The argument is very much the same as in the case of the
bosonic mass matrices. It relies on the fact that, independently of the 
vacuum $({\cal Q},\iota)$, the corresponding vacuum section reads
${\cal V}(x)=[(\iota(q),{\bf z}_0)]|_{\mbox{\tiny$q\in\pi_{\rm Q}^{-1}(x)$}}$.
Thus, the spectrum of the fermionic mass matrix is independent of $x\in\mm$. 
Of course, if two minima ${\bf z}_0,{\bf z}'_0$ of a given general Higgs potential
$V_{\mbox{\tiny H}}$ are on the same orbit, then the corresponding vacua 
are equivalent. As a result, the spectrum of ${\rm m}_{\mbox{\tiny F}}^2$ 
only depends on the orbit of some minimum. By the very construction of 
the fermionic mass matrix we have $[{\cal D},\rho_{\mbox{\tiny F}}(h)]=0$ 
for all $h\in{\cal H}$, such that ${\rm H}\simeq{\rm I}({\bf z}_0)$. Since 
${\rm m}_{\mbox{\tiny F}}\in\Omega^0(\mm,{\rm End}^-(\ep))$ is constant
on the reduced fermion bundle one can decompose the latter with 
respect to the eigenbundles of 
${\rm M}_{\mbox{\tiny F}}^\dagger{\rm M}_{\mbox{\tiny F}}$ and 
${\rm M}_{\mbox{\tiny F}}{\rm M}_{\mbox{\tiny F}}^\dagger$.\hfill$\Box$\\

For fixed ${\rm m}^2\in{\rm spec}({\rm m}_{\mbox{\tiny F}}^2)$ the Clifford 
module bundle $\xi_{\mbox{\tiny F}}{\mbox{\tiny$({\rm m}^2)$}}$ is regarded 
as the geometrical analogue of an {\it almost free fermion of mass m}. Here, 
``almost'' refers to the circumstance that neither the connection $\Theta$, 
nor the reduced representation $\rho_{\mbox{\tiny F,red}}$ is trivial, in 
general. Therefore, a non-trivial vacuum together with the topology of 
space-time may give rise to a non-trivial holonomy group analogously to 
the well-known Aharonov-Bohm effect. Notice that, if the spectrum of
the fermionic mass matrix is non-degenerated, then $\xi_{\mbox{\tiny F,red}}$ 
decomposes into the Whitney sum of the tensor product of the spinor bundle 
and appropriate Hermitian line bundles.\\

In what follows we will rewrite a general Dirac-Yukawa operator in terms of
a ``fluctuation'' of the vacuum at hand. For this let us call to mind the definition 
of the latter (c.f. \cite{tolksdorf'02}).\\

Let $(\Theta,{\cal V})$ be a vacuum pair that spontaneously breaks a
Yang-Mills-Higgs gauge theory that is defined by the data
$({\cal P}(\mm,{\rm G}),\rho_{\mbox{\tiny H}},V_{\mbox{\tiny H}})$. We
call a one-parameter family of Yang-Mills-Higgs pairs $(A_t,\varphi_t)\in
{\cal A}(\xi_{\mbox{\tiny H}})\times\Gamma(\xi_{\mbox{\tiny H}})$
($0\leq t\leq 1$) a {\it fluctuation of the vacuum} if there is a
Yang-Mills-Higgs pair $(A,\varphi)$, such that 
$A_t = \Theta + t(A-\Theta)$ and $\varphi_t = {\cal V} + t\varphi$, where
$A$ is supposed to be associated to a non-reducible connection on 
${\cal P}(\mm,{\rm G})$ and $\varphi$ is supposed to be in the
``unitary gauge'', i.e. $\iota^*\varphi\in\Gamma(\xi_{\mbox{\tiny H,phys}})$.
Here, we make use of the fact that the reduced Higgs bundle, when 
considered as a real vector bundle, decomposes into the Whitney sum
\bb
\xi_{\mbox{\tiny H,red}}=\xi_{\mbox{\tiny G}}\op\xi_{\mbox{\tiny H,phys}}
\ee 
of two real sub-vector bundles representing the Goldstone and the physical 
Higgs boson (c.f. loc sit).\\ 

By identifying a connection with its connection form we may 
write a fluctuation of the canonical Dirac-Yukawa operator 
$\ddd_{\!\mbox{\tiny${\cal D}$}}$ as follows: 
\bb
\label{fluc dir-yuk op}
D_{\mbox{\tiny Y,t}} := \ddd_{\!\mbox{\tiny${\cal D}$}} +
t\,{/\!\!\!\!A_{\mbox{\tiny fl}}},
\ee
where the ``fluctuation'' reads: ${/\!\!\!\!A_{\mbox{\tiny fl}}}:=
\gamma(A-\Theta) + \gamma_5\ot G_{\mbox{\tiny Y}}(\varphi)$. 
We stress that the zero order operator 
$D_{\mbox{\tiny Y}} - \ddd_{\!\mbox{\tiny${\cal D}$}}$ defines a 
fluctuation of the vacuum iff the unitary gauge exists. This holds true, e.g., 
in the case of rotationally symmetric Higgs potentials, like in the (minimal) 
Standard Model (see again loc sit). Therefore, every Dirac-Yukawa operator 
on the fermion bundle can be regarded as a fluctuation of the canonical 
Dirac-Yukawa operator on the reduced fermion bundle, provided that the 
unitary gauge exists. Notice that either of the two terms on the right 
hand side of (\ref{fluc dir-yuk op}) transform gauge covariantly with
respect to the invariance group of the vacuum. The sum of both, however,
is covariant with respect to the original gauge group ${\cal G}$. In other
words: the fluctuation ${/\!\!\!\!A_{\mbox{\tiny fl}}}$ of the vacuum makes 
the canonical Dirac-Yukawa operator $\ddd_{\!\mbox{\tiny${\cal D}$}}$ on
the (reduced) fermion bundle also ${\cal G}-$covariant.\\ 

Since the Dirac-Yukawa operator 
\bb
i\ddd_{\!\mbox{\tiny${\cal D}$}}\equiv i\ddd-{\rm m}_{\mbox{\tiny F}}
\ee 
is the geometrical analogue of Dirac's original first order operator
$i\ddd-{\rm m}$, the fluctuation (\ref{fluc dir-yuk op}) might be regarded as 
a geometrical variant of what is usually referred to as ``minimal coupling''. In 
the case at hand, however, the minimal coupling (\ref{fluc dir-yuk op}) 
naturally includes the gravitational field and the states of the Higgs boson.
We stress that on the basis of general relativity (without an a priori 
cosmological constant) it would be inconsistent to assume a non-trivial
fermionic mass matrix together with a trivial gravitational field\footnote{Of 
course, this should not be confounded with the common assumption, e.g., in 
particle physics, of a {\it negligible} contribution of the gravitational 
field.}. 
Thus, a non-trivial ground state of the Higgs boson yields a non-trivial 
gravitational field, in general.\\

In this section we have presented a physically intuitive interpretation of
a geometrically distinguished class of Dirac-type operators\footnote{For 
instance, Dirac operators of simple type are fully characterized by their 
Bochner-Lichnerowicz-Weitzenb\"ock decomposition, see, e.g., 
\cite{ackerm/tolksd'96}. Moreover, as already mentioned they constitute the 
biggest class of Dirac-type operators such that their connection classes have a 
canonical representative.}. This interpretation in turn permits a different 
geometrical interpretation of minimal coupling with the basic feature 
of including the Higgs boson and the gravitational field. From a geometrical 
perspective the back and forth of both interpretations may be most 
transparently summarized by the canonical isomorphism between the 
fermion bundle and the reduced fermion bundle and the fact that the latter 
decomposes into the eigenbundles of the fermionic mass matrix. Of course, 
these identifications depend on the vacuum (pair) only up to gauge equivalence.\\

So far the three datasets $(\mm, g_{\mbox{\tiny M}})$,
$({\cal P}(\mm,{\rm G}),\rho_{\mbox{\tiny H}},V_{\mbox{\tiny H}})$ and
$({\cal P}(\mm,{\rm G}),\rho_{\mbox{\tiny F}},{\cal S})$ have been assumed
to be given. Moreover, these sets are connected only by a Yukawa mapping
(\ref{yukawa mapping}). In our final section we want to indicate how to connect 
these sets by postulating a ``universal Lagrangian density'' that is naturally 
defined on ${\cal D}(\xi_{\mbox{\tiny F}})$. 

\section{Dirac Potentials and Lagrangians}
In this paper we consider a fermion as a geometrical object that is defined 
by the data $({\cal P}(\mm,{\rm G}),\rho_{\mbox{\tiny F}},{\cal S})$, where 
$(\mm,g_{\mbox{\tiny M}})$ is supposed to be given. With respect to this
setup there is a distinguished class of first order differential operators
acting on the states of the fermion. As an additional input we considered the
data $({\cal P}(\mm,{\rm G}),\rho_{\mbox{\tiny H}},V_{\mbox{\tiny H}})$ that
geometrically define a Yang-Mills-Higgs gauge theory. In order to combine
both datasets we introduced the Yukawa mapping and thereby a specific
class of Dirac operators of simple type called general Dirac-Yukawa operators.
In fact, the Yukawa mapping generalizes what is known in physics as 
``Yukawa coupling''. If the Yang-Mills-Higgs gauge theory is spontaneously
broken, then the fermion decomposes into almost free fermions. Each of these
fermions is geometrically represented by a non-flat Clifford module bundle,
where the curvature is determined by the gravitational field together with 
the mass of the free fermion in question. However, since the (pseudo) metric
structure has been fixed right from the beginning, these two contributions
to the total curvature of the fermion bundle are thus far independent of 
each other. Of course, when seen from a physical perspective, this seems 
unsatisfying. One may expect that the masses of the fermions give a 
contribution to the gravitational field. The most natural way to achieve this 
is the following construction, which naturally incorporates the dynamics of the
gravitational field in the geometrical picture presented here. For this we 
introduce the following {\it universal Lagrangian mapping}:
\bb
\label{canonical lagrangian}
{\cal L}:\, {\cal D}(\xi_{\mbox{\tiny F}})&\longrightarrow&\Omega^{2n}(\mm)\cr
D &\mapsto& \ast{\rm tr}V_{\mbox{\tiny D}}.
\ee
We call the zero order operator $V_{\mbox{\tiny D}}:=
D^2-\triangle_{\mbox{\tiny D}}\in\Omega^0(\mm,{\rm End}(\ep))$ 
the {\it Dirac potential} associated with $D$. It is fully 
determined by the Dirac-type operator in
question and can explicitly be calculated, for instance, by using 
the generalized Bochner-Lichnerowicz-Weitzenb\"ock decomposition 
formula (see Eq. 3.13 in \cite{ackerm/tolksd'96}). We mention that the Dirac 
potential generalizes the Higgs potential, for it can be shown that (at least 
with respect to the Euclidean signature) the Lagrangian of a Yang-Mills-Higgs
gauge theory can be recovered from an appropriate Dirac-type operator 
(see \cite{tolksdorf'98}). In this case, however, one has to take into account 
anti-particles as well (see \cite{tolksdorf'01}).\\ 

In the case of the canonical Dirac-Yukawa operator  
$\ddd_{\!\mbox{\tiny${\cal D}$}}$ one obtains the following Lagrangian density:
\bb
{\cal L}(\ddd_{\!\mbox{\tiny${\cal D}$}}) = 
\ast{\rm tr}({\mbox{\small$\frac{r_{\rm M}}{4}$}} + {\rm m}_{\mbox{\tiny F}}^2).
\ee
Here, $r_{\mbox{\tiny M}}\in{\cal C}^\infty(\mm)$ is the scalar curvature
with respect to an appropriate O(2n)-reduction of the (oriented) frame bundle 
of $\mm$. As a consequence, space-time must be an Einstein manifold, where the 
gravitational field is now dynamically determined by the masses of the 
almost free fermions. Moreover, the mass of an almost free fermion also 
determines its curvature. In other words: in the ground state both the geometry
of space-time and of the (reduced) fermion bundle is determined by the 
fermionic masses. We summarize this by saying that the ``fermionic
vacuum'' gives rise to a Lagrangian of the form
\bb
\label{fermionic vacuum}
{\cal L}(\ddd_{\!\mbox{\tiny${\cal D}$}}) &\sim&
{\mbox{\small$<\!{\rm m}^2_{\mbox{\tiny F}}\!>$}}\,\mu_{\mbox{\tiny M}},
\ee
with ${\mbox{\small$<\!{\rm m}^2_{\mbox{\tiny F}}\!>$}}:=
{\mbox{\small$\frac{1}{{\rm N}_{\rm F}}$}}
\sum_{k=1}^{\mbox{\tiny${\rm N}_{\rm F}$}}{\rm m}_k^2$ and
$\mu_{\mbox{\tiny M}}$ the appropriate volume form determined by 
the fermionic mass.\\ 

Of course, since the Lagrangian (\ref{fermionic vacuum})
is fully determined by the spectrum of the fermionc mass matrix, 
it is invariant with respect to both the gauge group of 
general relativity (i.e. the group of volume preserving automorphisms of the 
oriented frame bundle) and to the invariance group of the vacuum. Moreover, 
it only depends on the orbit of some minimum and not of the vacua with 
respect to this minimum. Therefore, the Lagrangian of the fermionic vacuum 
is indeed a spectral invariant.

\section{Summary and Outlook}
We have presented a geometrical setup permitting a geometrical interpretation 
of the fermionic masses. Moreover, we have shown how the fermionic mass 
determines the geometry of space-time and that of the Clifford module bundle 
which geometrically represents almost free fermions. The fact that the 
spectral invariant 
(\ref{fermionic vacuum}) of the fermionic vacuum is proportional to the 
mean value of the fermionic masses is clearly due to the circumstance 
that the fermion bundle breaks into a Whitney sum with respect to any given 
vacuum. This splitting geometrically describes what is usually referred to 
as ``particle multiplet''. With respect to a ``linear fluctuation'' of a vacuum 
(pair) the canonical Dirac-Yukawa operator becomes covariant with 
respect to the full gauge group. However, in this case the corresponding 
canonical Lagrangian determines neither the dynamics of the Higgs
boson, nor that of the gauge boson. Moreover, the appropriate Lagrangian
reduces to that of the fermionic vacuum. For this it might be more natural 
to consider a ``quadratic fluctuation'' of a vacuum. As we have mentioned
before, in this case the Lagrangian mapping yields (up to a constant) the 
bosonic Lagrangian of the Standard Model with gravity included. 
Notice that for ${\rm dim}(\mm)=4$ there are no 
``higher fluctuations'' of the vacuum. Moreover, by geometrically 
incorporating the notion of anti-particles (i.e. a real structure) the quadratic 
fluctuations give rise to the same dynamics for the fermions as the linear 
fluctuations of the vacuum do. This has been discussed, e.g, in \cite{tolksdorf'01} 
for the case of the Euclidean signature. The main reason for using the Euclidean 
signature was that we are dealing with a universal action instead of a Lagrangian. 
However, when gravity is taken into account, the latter seems more appropriate 
since a Lagrangian is a density and thus a purely local object. And because it is a
density, the signature of $g_{\mbox{\tiny M}}$ does not matter. Moreover, 
Clifford module bundles always refer to some Clifford bundle over $\mm$. But
this in turn obviously refers to some chosen O(2n)-reduction of the frame 
bundle of $\mm$, i.e. to some fixed $g_{\mbox{\tiny M}}$. However, when 
$g_{\mbox{\tiny M}}$ is physically interpreted as a gravitational field, it cannot
be fixed a priori, for it has to satisfy, e.g., Einstein's equation. This is obviously 
a dilemma one always has to face in if gravity is taken into account. The 
philosophy of the paper at hand with respect to the Lagrangian mapping 
(\ref{canonical lagrangian}) is as follows: the field equations determined by the 
corresponding Lagrangian are considered as ``constraints'' of how to glue 
together the local pieces to give rise to global geometrical objects like, e.g., the 
fermion bundle. We consider this interpretation of the Euler-Lagrange 
equations to hold true, especially in the case of the Einstein equation.\\

Interestingly, there are certain parallels between the geometrical setup 
presented here and what is called ``almost commutative models'' in the 
literature. In particular, the canonical Dirac-Yukawa operator corresponds 
to the ``total Dirac operator'' and ${\cal D}$ to the ``internal Dirac operator'' 
in the Connes-Lott description of the Standard Model within the frame of 
A. Connes' non-commutative geometry (see, e.g., 
\cite{connes'94} or \cite{gra et al'02}). Like in the case of the Connes-Lott 
model one also has a ``fermion doubling'' in the geometrical frame presented 
here. This still has to be carefully analyzed, for we can perhaps work
with the physical signature of $g_{\mbox{\tiny M}}$. Concerning quantization
it seems challenging to try to understand what it geometrically means to 
``quantize'' the above mentioned constraints. This, of course, is still an open
question and has not been addressed in this paper. Instead, the main objective
here was to explore the geometrical meaning of the fermionic ``mass 
without mass''.

\vspace{0,8cm}

\noindent
{\bf Acknowledgments}\\
I would like to thank E. Binz for very interesting and stimulating discussions.

\end{document}